\documentclass{svjour3}                     % onecolumn (standard format)
\smartqed  % flush right qed marks, e.g. at end of proof
\usepackage{graphicx}
\begin{document}

\title{{Search for exoplanets in M31 with pixel-lensing 
and the PA-99-N2 event revisited}}
\author{Gabriele Ingrosso  \and 
 Sebastiano Calchi$~$Novati    \and
 Francesco De Paolis         \and
 Philippe Jetzer           \and
Achille Nucita           \and
Alexander Zakharov}

\authorrunning{Ingrosso et al.} % if too long for running head

\institute{
Gabriele Ingrosso \at 
Dipartimento di Fisica, Universit\`a del Salento, and
{\it INFN} Sezione di Lecce,  I-73100 Lecce, Italy, 
\email{ingrosso@le.infn.it}           %  \\
\and 
Sebastiano Calchi Novati \at 
Dipartimento di Fisica, Universit\`{a} di Salerno, and
{\it INFN} Sezione di Napoli, I-84081 Baronissi, Italy,
\email{novati@sa.infn.it}           %  \\
\and
Francesco De Paolis \at 
Dipartimento di Fisica, Universit\`a del Salento and
{\it INFN} Sezione di Lecce, CP 193, I-73100 Lecce, Italy,
\email{depaolis@le.infn.it}           %  \\
\and
Philippe Jetzer
\at Institute for Theoretical Physics,
University of  Z\"{u}rich, Winterthurerstrasse 190,
CH-8057 Z\"{u}rich, Switzerland, \email{jetzer@iftp.uzh.ch}   
\and 
Achille Nucita \at
XMM-Newton Science Operations Centre, ESAC, ESA,
PO Box 50727, 28080 Madrid, Spain, 
\email{achille.nucita@sciops.esa.int} 
\and 
Alexander Zakharov \at 
Institute of Theoretical and Experimental Physics,
B. Cheremushkinskaya 25, 117259 Moscow, and 
Bogoliubov Laboratory of Theoretical Physics, 
Joint Institute for Nuclear Research, 141980 Dubna, Russia,
\email{zakharov@itep.ru} }

\date{Received: date / Accepted: date}

\maketitle

\begin{abstract}
Several exoplanets have been detected towards the Galactic bulge 
with the microlensing technique. We show that exoplanets in M31 may also be 
detected with the pixel-lensing method, if telescopes making high
cadence observations of an ongoing microlensing event are used.  
Using a Monte Carlo approach we find that the mean mass 
for detectable planetary systems is about $2~M_{\rm {J}}$. However, 
even small mass exoplanets ($M_{\rm P} < 20~M_{\oplus}$) can cause significant 
deviations, which are observable with large telescopes.
We reanalysed the POINT-AGAPE
microlensing event PA-99-N2. 
First, we test the robustness
of the binary lens conclusion for this light curve.
Second, we show that for such long duration and bright
microlensing events, the efficiency for finding
planetary-like deviations is strongly enhanced  with respect to that
evaluated for all planetary detectable events.


%95.75.De Photography and photometry (including microlensing techniques)
%97.82.-j Extrasolar planetary systemsPACS 
%98.62.Sb Gravitational lenses and luminous arcs
%98.56.Ne Spiral galaxies (M31 and M33)

\end{abstract}

\section{Introduction}
\label{intro}

Gravitational microlensing technique initially developed to search for 
MACHOs in the Galactic halo can be used to infer the presence of exoplanets 
around lens stars \cite{gould08,bennett09}. Indeed, the planet
orbiting the primary lens star may induce significant 
perturbations to the standard (single lens) 
Paczy\'{n}ski like microlensing light curves \cite{maopacz91,pacz96}. 
Until now, 10 exoplanets has been detected towards the Galactic bulge
(see http://exoplanet.eu)
and the least massive planets have masses of about 3, 5 and 13 $M_{\oplus}$.
The planet orbital 
separations are in the range 2--5 AU (about the Einstein ring radius).
Microlensing technique complements the planet detections by 
 other  methods
(Doppler radial velocity measurements and transits are among the most 
efficient) 
\footnote{At the moment about 400 exoplanets have been discovered, 
basically at relatively small distances from the Earth 
(see http://exoplanet.eu).} 
that are more sensitive to large planet masses (Jupiter-like planets) 
at small orbital distances.
Microlensing gives also the opportunity to detect planets lying in M31 
\cite{covone,baltzgondolo2001}.
In this case, however, the source stars are not resolved by ground based
telescopes,  a situation referred to as ``pixel-lensing''
\cite{crotts92,agape93,gould96,agape97}. 
Until now, however, only 25 microlensing events have been observed 
towards M31 \cite{review_novati} and in one case a deviation of the 
microlensing light curve from the standard Paczy\'{n}ski shape has been 
observed and attributed to a secondary component orbiting the lens star 
\cite{an04}. However, new observational campaigns towards M31 
have been undertaken \cite{kerins06,seitz08,novati09}
and hopefully new exoplanets might be detected in the future. 

The possibility to detect planets in pixel-lensing observations towards M31 
has been already explored 
\cite{covone,chungetal06,kimetal07}. 
The analysis for planet detection, however, has been performed by using
a fixed configuration of the underlying Paczy\'{n}ski light curve.
In the present work, using a Monte Carlo (MC) approach 
\cite{mnras} we explore chances to 
detect exoplanets in M31, by considering the multi-dimensional 
space of parameters for both lensing and planetary systems. 
By using the method of residuals,
we select the light curves that show detectable deviations with 
respect to the Paczy\'{n}ski shape.
The advantage of the MC approach is that of allowing
a detailed characterisation of the sample of microlensing events 
for which the planetary deviations are more likely to be detected.

\section{Generation of Planetary Microlensing Events}
\label{sec:1}

Assuming that the lens is a planetary system lying in the M31 bulge or disk 
we generate a sample of pixel-lensing events towards M31. 
Microlensing parameters are selected according to the method outlined in 
\cite{kerins01,ingrosso06,ingrosso07}. To select planetary parameters 
(mass $M_{\rm P}$ and 
orbital period $P$) we adopt the relation \cite{tremaine}
\begin{equation}
   dn(M_{\rm P},P) \propto ~ M_{\rm P}^{-\alpha}~ P^{-\beta}~
   \left(\frac {dM_{\rm P}}{M_{\rm P}}\right) ~ \left( \frac {dP}{P}\right)~~,
\end{equation}
with $\alpha=0.11$ and $\beta=-0.27$.
This relation was obtained by investigating the distribution of masses
and orbital periods of 72 exoplanets 
(detected  mainly by Doppler radial velocity method), 
and taking into account the selection
effects caused by the limited velocity precision and duration of existing 
surveys 
\footnote{A recent re-analysis \cite{jiangetal} of the mass-period relation 
employing a larger catalog of 175 exoplanets detected by the 
Doppler radial velocity method is consistent with that of \cite{tremaine}.}. 
The upper limit of the planetary mass is set at $10~M_{\rm {J}}$
which roughly corresponds to the lower mass limit for brown dwarfs.
Moreover, following the numerical simulations for the planetary 
system formation \cite{ida}, we set the lower planetary mass limit at 
$M_{\rm P}=0.1~M_{\oplus}$.  
Once the masses of the binary components and the planet period have been 
selected, the binary separation $d_{\rm P}$ 
is obtained by assuming a circular orbit.
We also suppose that planetary perturbation time scales are much shorter
than orbital periods of planets. 
All distances are normalised to the Einstein radius
$R_{\rm E}$ of the total mass $M$ of the lens system 
\begin{equation}
R_{\rm E} = \sqrt{ \frac{4GM}{c^2} 
\frac {D_{\rm S}-D_{\rm L}} {D_{\rm S}}}~,
\end{equation}
where $D_{\rm S}$ ($D_{\rm L}$) is  the source (lens) distance.
Moreover, it is assumed that all stars in M31 have planets.

%We consider different lines of sight to M31, observational conditions and 
%sampling times of the light curves. 
We simulate pixel-lensing light curves
assuming a regular sampling (between $2$ and $24$ hours), telescopes of 
different diameters in the range $2-8$ m, typical observational 
conditions and an exposure time $t_{\rm exp}$ of 30 and 60 min. 
Moreover, since in pixel-lensing towards M31 the bulk of 
the source stars are red giants (with large radii), we take into 
account the source finiteness by averaging the planetary magnification
$A_{\rm P}(t_i)$ (numerically evaluated by using the binary lens equation 
\cite{witt90,witt95}) on the source size. 
A microlensing light curve is given by
\begin{equation}
S_{\rm P}(t_i) = f_{\rm bl} + f_0 [\langle A_{\rm P}(t_i) \rangle -1]~,
\end{equation}
where $f_{\rm bl}$ is the  background signal from M31 and the sky,
$f_0$ is the unamplified source star flux and $ \langle A_{\rm P}(t_i) \rangle$
 the time 
varying magnification, that is averaged on the (projected) source size $\Sigma$
\begin{equation}
\langle A_{\rm P}(t_i) \rangle = \frac {\int_0^{2\pi} d\theta \int_0^{\rho}
A_{\rm P}(\tilde{\theta},\tilde{\rho};t_i)I(\tilde{\rho})
\tilde{\rho}d\tilde{\rho}} { 2 \pi \int_0^{\rho} I(\tilde{\rho})
\tilde{\rho}d\tilde{\rho}} ~, \label{finite1}
\end{equation}
where $\rho = {\theta_{\rm S}}/{\theta_{\rm E}}$
is the normalised angular size of the source
($\theta_{\rm S} = R_{\rm S}/D_{\rm S}$, $R_{\rm S}$ being the source radius,
 $\theta_{\rm E} = R_{\rm E}/D_{\rm L}$ is the angular Einstein radius)
and $I(\tilde{\rho})$ the intensity profile of the source including limb 
darkening (see \cite{mnras} for more details).

Planetary perturbations in the light curves occur when the source star 
trajectory in the lens-plane (the plane orthogonal to the line of sight to the
M31 star source, passing at the lens position) crosses and/or passes near 
caustics. This is the caustic set of the source positions at which the 
magnification is infinite in the ideal case of a point source. Clearly
for realistic sources of finite size the magnification gets still large, but
finite.

Light curves that show significant ($>3 \sigma$) flux variations 
with respect to the background noise in at least three points   
identify a sample of ``detectable'' events. These curves are fitted 
with a standard Paczy\'{n}ski form 
\begin{equation}
   S^0(t) = f_{\rm bl}^0+f_0^0 [ \langle A^0(t_i) \rangle -1]~,
\end{equation}
where $ \langle A^0(t_i) \rangle$ 
is the usual magnification for a single lens and 
the averaging is made on the projected source size $\rho$.

\begin{figure}
\includegraphics[width=0.75\textwidth]{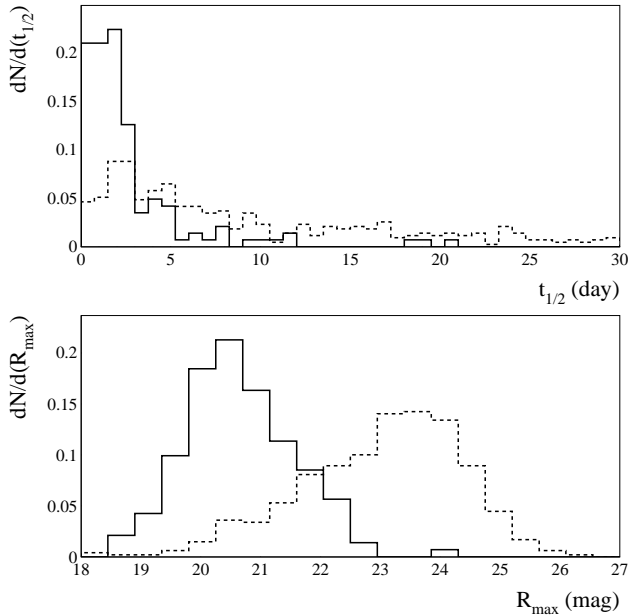}
\caption{Normalised (to unity) 
 distributions of $t_{1/2}$ (top panel) and $R_{\rm max}$
(bottom panel)  for planetary selected events with $\chi_{\rm r} > 4$, 
$N_{\rm good} > 3$ and $\langle\epsilon\rangle_{\rm max}>0.1$. 
Here and in the following figures we select the following parameters:
telescope diameter D=8 m, an exposure time $t_{\rm exp} = 30$ min, a sampling 
time of the light curve of 2 hours. Solid lines correspond to I class events,
dashed lines to II class events.}
\label{proceeding1}       % Give a unique label
\end{figure}
%
%

\begin{figure}
%\includegraphics{proceeding2.eps}
\includegraphics[width=0.75\textwidth]{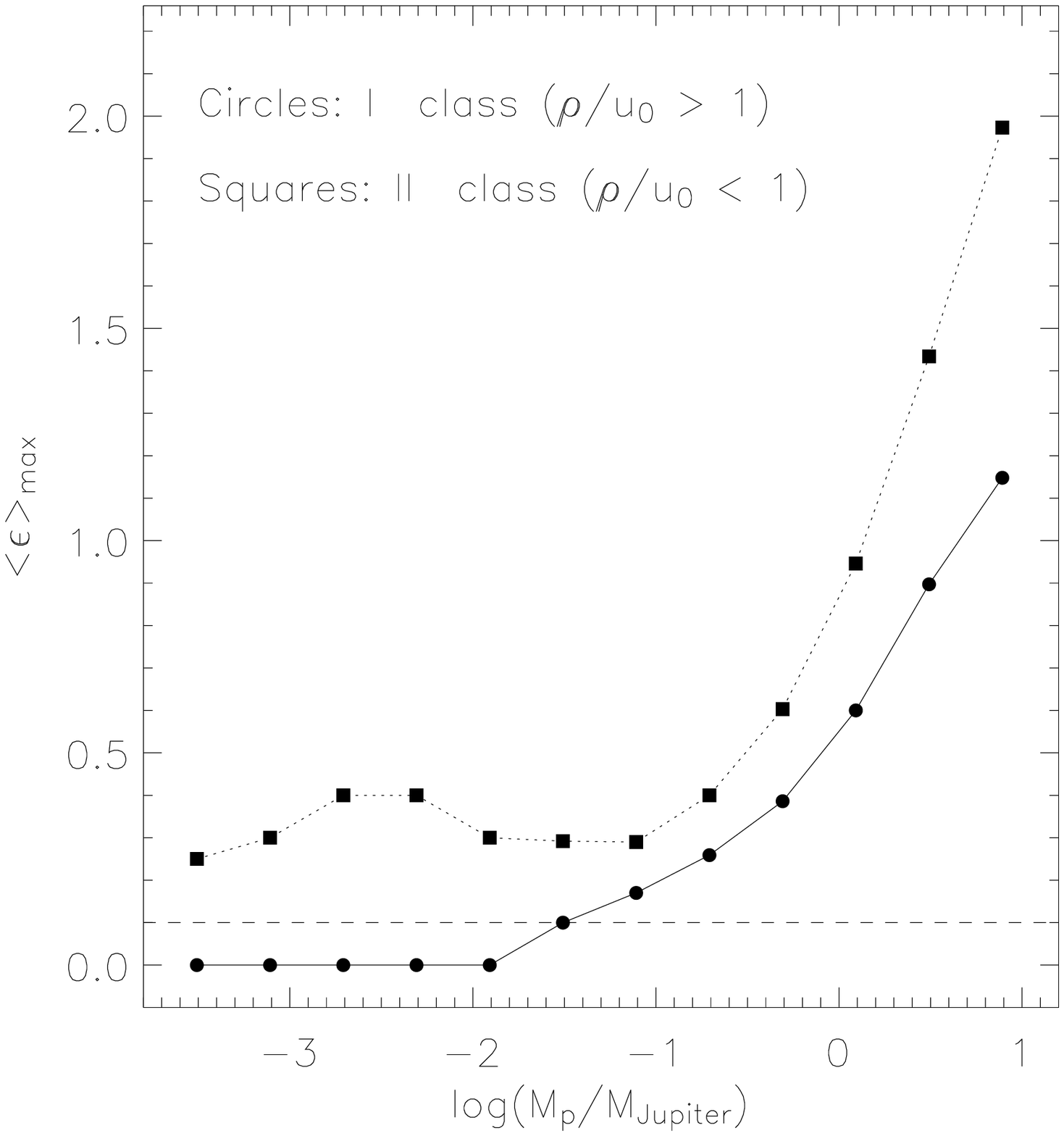}
\caption{The maximum relative planetary magnification 
$ \langle \epsilon \rangle_{\rm max}$ 
is given as a function of $M_{\rm P}$.}
\label{proceeding2}       % Give a unique label
\end{figure}

Next, to select light curves with detectable planetary features  
we consider two indicators:
{\it (i)} the overall deviation (in units of error-bars) of the 
planetary light curves from the corresponding Paczy\'{n}ski best fit, 
as measured by the sum of residuals  
$\chi_{\rm r}^2 = \sum_i \chi_{\rm r}^2(t_i)/N_{\rm tot}$ 
($N_{\rm tot}$ is the total number of points) and {\it (ii)} the 
number of points $N_{\rm good}$, even not consecutive, which
deviate significantly from the Paczy\'{n}ski fit. 
By a direct survey of many light curves we fix the thresholds 
$\chi_{\rm r~th}=4$ and $N_{\rm good~th}=3$ 
and we select a sub-sample of light curves 
with detectable planetary features having $\chi_{\rm r}^2 > 4$ and 
$N_{\rm good}>3$.
\begin{figure}
\includegraphics[width=0.75\textwidth]{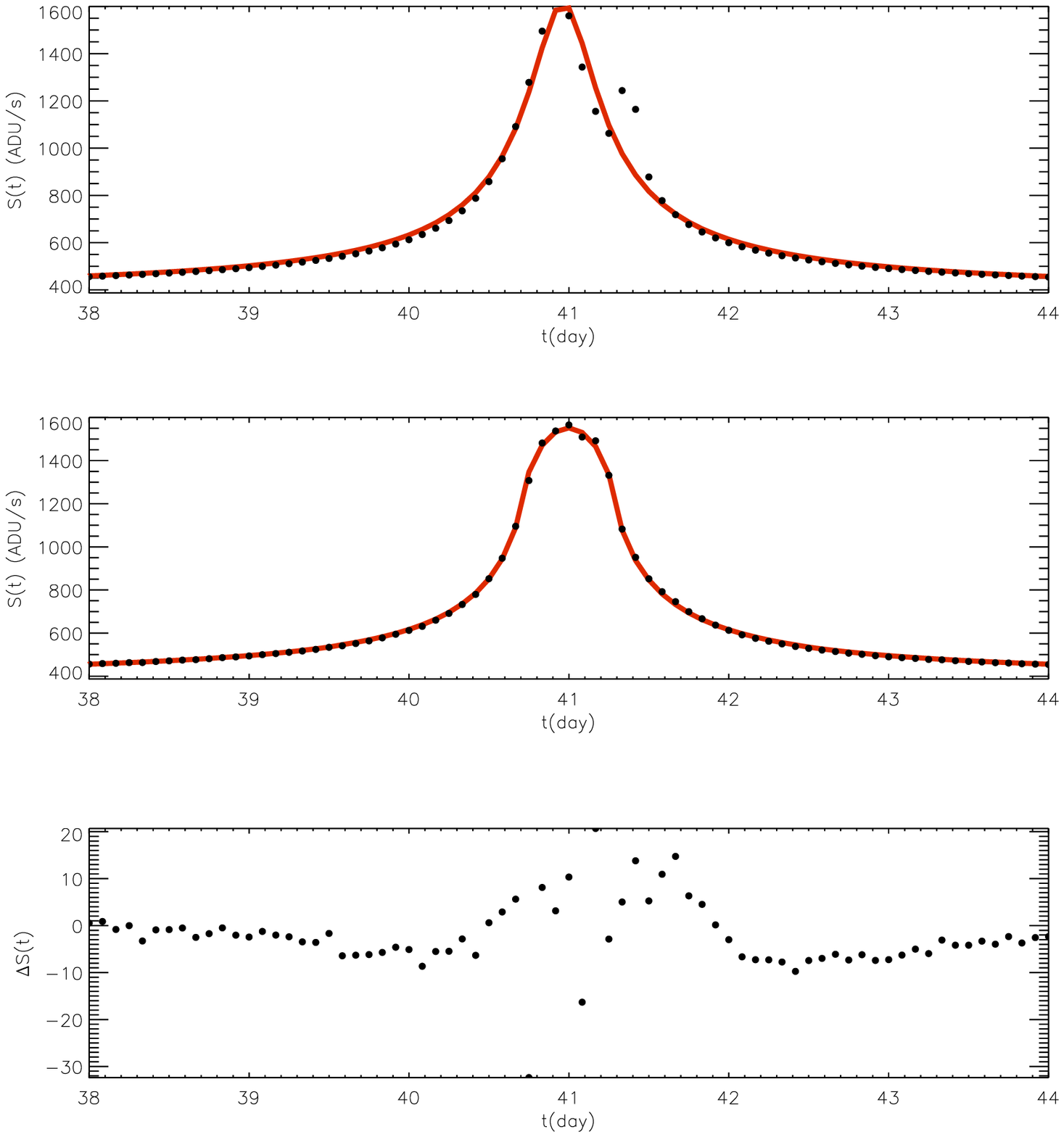}
%  \includegraphics{proceeding3.eps}
\caption{Simulated I class event (points) and Paczy\`nki fit (solid line).
Some parameter values: 
$\rho/u_0=2.14$, 
$R_{\rm max}=19.0$ mag, 
$t_{1/2}=0.6$ day, 
$M_{\rm P}=1.04~M_{\rm {J}}$,
$M =0.55~M_{\odot}$ ($M$ is the total mass of the lens system),
$d_{\rm P}=3.6$AU,
$R_{\rm E}=5.55$AU, 
$\chi_{\rm r}=46$, 
$ \langle \epsilon\rangle_{\rm max}=0.2$.}
\label{proceeding3}       % Give a unique label
\end{figure}
However, in this way, we find that some of the selected light curves show 
overall distortions (with respect to the Paczy\'{n}ski shape) which hardly 
could be attributed in real observations to a planet orbiting the primary lens
star (we mean that it is not easy to reconstruct parameters of planetary 
systems from such observational data). 
This happens, in particular, for events with small planetary mass and/or 
large (projected) source radius, for which the finite source effects 
are maximised.
A measure of the relevance of these effects is given by  
the relative (with respect to the Paczy\'{n}ski value) planetary magnification 
averaged on the source area
\begin{equation}
 \langle \epsilon(t_i)\rangle = \left( \frac 
{\int_{\Sigma} d^2 \vec{x} ~ 
[|A_{\rm P}(\vec{x},t_i)-A^0(\vec{x},t_i)|/A^0(\vec{x},t_i)]}
{\int_{\Sigma} d^2 \vec{x}} \right) ~.
\end{equation}
To be conservative, events with maximal finite source effects
are eliminated from the following analysis by requiring, 
besides conditions {\it (i)} and {\it (ii)},
 that  {\it (iii)} $ \langle \epsilon\rangle_{\rm max} > 0.1$, where 
$ \langle \epsilon \rangle_{\rm max}$ 
is the maximum value of $ \langle \epsilon(t_i)\rangle$ 
along the whole light curve.
Conditions {\it (i)} and {\it (ii)} are particularly efficient to select light 
curves with a large number of points deviating from the Paczy\'{n}ski best 
fit, the condition {\it (iii)} ensures the presence on the light curve of at 
least one clear planetary feature.

%small ($u_0<<1$) impact parameter (large amplified events for 
%which the planetary deviations are caused by intersection of the source 
%trajectory with the central caustic) and large (relative) projected source 
%radius ($\rho >> u_0$).

\section{Results}
\label{sec:2}
\begin{figure}
\includegraphics[width=0.75\textwidth]{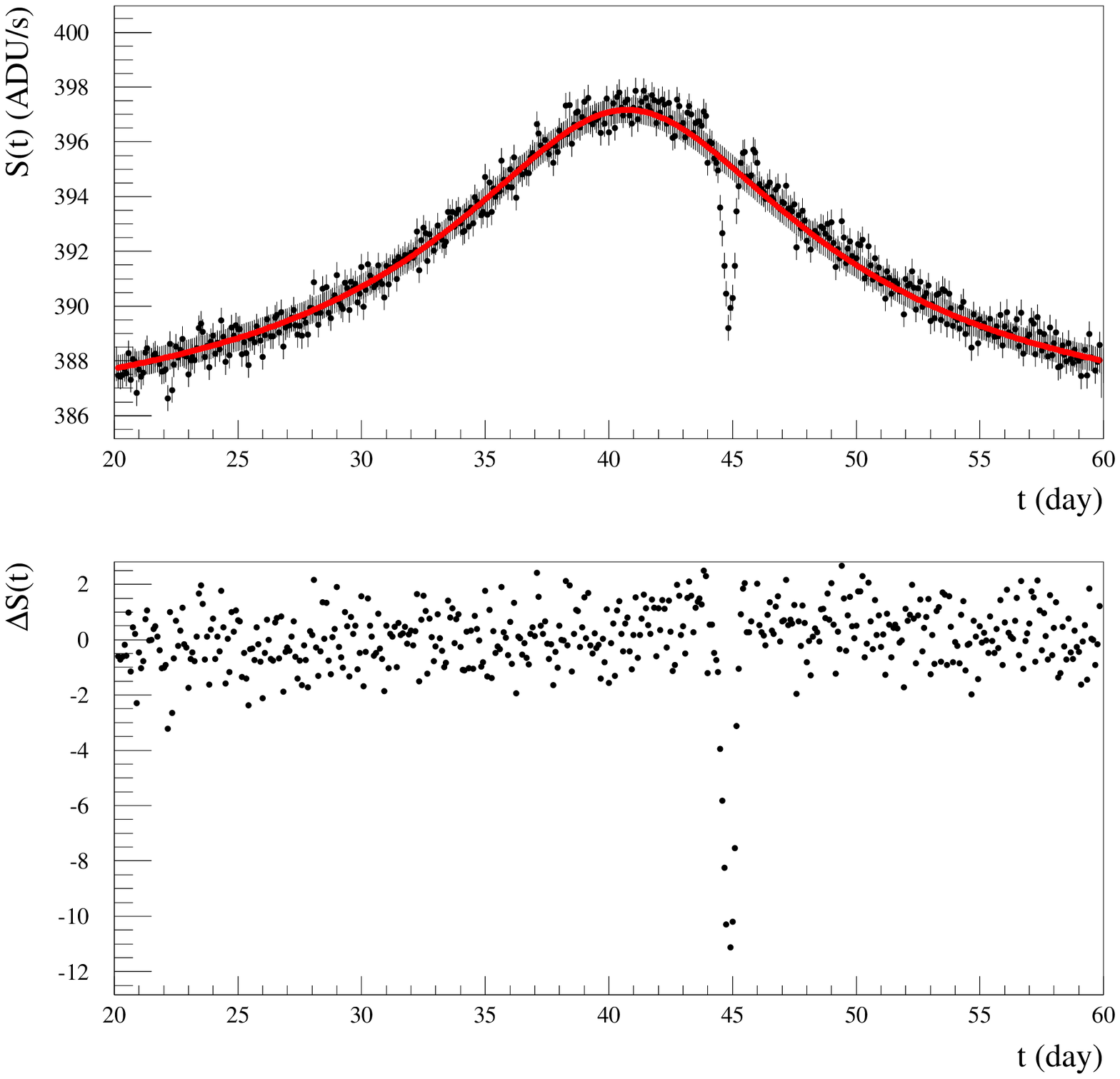}
\caption{Simulated II class event (points) and Paczy\`nki fit 
(thin solid line).
Some parameter values: 
$\rho/u_0=0.04$, 
$R_{\rm max}=24.0$ mag, 
$t_{1/2}=18.7$ day, 
$M_{\rm P}=0.3~M_{\oplus}$,
$M =0.31~M_{\odot}$ ($M$ is the total mass of the lens system),
$d_{\rm P}=3.8$AU, 
$R_{\rm E}=3.3$AU, 
$\chi_{\rm r}=8.0$, 
$ \langle \epsilon\rangle_{\rm max}=0.4$.}
\label{proceeding4}       % Give a unique label
\end{figure}

In Fig. \ref{proceeding1} (for $D=8$ m), 
we give the distributions of $t_{1/2}$ and $R_{\rm max}$ for the
events with detectable planetary deviations 
($\chi_{\rm r}^2 > 4$, 
$N_{\rm good}>3$ and $ \langle \epsilon \rangle_{\rm max} > 0.1$).
Here, $t_{1/2}$ is the full-width half-maximum microlensing
event duration  $t_{1/2} = t_\mathrm{E} ~f(u_0)$, where $f(u_0)$ is a 
function \cite{gould96} of the dimensionless impact parameter $u_0$ 
and $R_{max}$ is the magnitude in the $R$-band corresponding 
to the flux variation at the maximal Paczy\'nski {\bf magnification }
$\Delta f^0_{~max} = f_0^0(A^0_{~max}-1)$.
%Here, $t_{1/2}$ is the the full-width half-maximum event duration
%and $R_{\rm max}$ is the magnitude in the $r$-band corresponding to the 
%maximum flux variation of the source. 
In the Fig. \ref{proceeding1}  we discriminate two classes of 
events (indicated by I and II), according to the ratio $\rho/u_0 >1$ (solid 
lines) or $\rho/u_0 <1$ (dashed lines). The events of the I class with 
$\rho/u_0 >1$ have short time durations ($ \langle t_{1/2}\rangle \simeq 1.6$ 
day) and 
larger flux variations ($ \langle R_{\rm max}\rangle \simeq 20.6$ mag). 
In these events planetary deviations are caused by the 
source trajectory crossing (in the lens plane) the central caustic region, 
close to the primary lens star.
The events of the II class, with $\rho/u_0 <1$, have longer durations 
($\langle t_{1/2} \rangle =6.4$ day) and smaller flux variations  
($\langle R_{\rm max} \rangle=23.1$ mag). Planetary perturbations 
in these cases are (mainly) 
caused by the intersection of the source trajectory with the planetary 
caustics and may also appear at times  
far from the maximum magnification time $t_0$.

The fraction of I class events is about 5\%. However, it turns out   
that the probability to have detectable planetary features in these events
(that however are rare) is higher. This happens since the crossing of the 
central caustic is more probable in I class events with $\rho/u_0>>1$.
On the contrary, the generated 
events of the II class are more numerous, but have 
a smaller probability to show detectable planetary features. 

The existence of the two classes of planetary events in pixel-lensing towards 
M31 is also evident in the Fig. \ref{proceeding2}, where we show 
$\langle \epsilon \rangle_{\rm max}$ 
(averaged on a large number of MC events of the 
same planetary mass $M_{\rm P}$) as a function of $M_{\rm P}$.  
We find that $\langle \epsilon \rangle_{\rm max}$ 
increases with increasing $M_{\rm P}$, a result that is expected 
since the size of the caustic region is increasing with the planetary mass 
\cite{gould92,Zakharov_Sazhin_UFN98,bozza99}. 
We also verify that the finite size effects are more important for I class 
events. Indeed, for a given mass $M_{\rm P}$, 
the events of the I class (with $\rho/u_0>1$) have smaller 
values of $\langle \epsilon \rangle_{\rm max}$ with respect to the 
corresponding II class events. 
 
An example of light curve for a I class event is shown in the central panel 
of the Fig. \ref{proceeding3}. Here, we can see how the planetary deviations 
in the light curve (upper panel) obtained by using the point-like source 
approximation are washed out when the magnification  is averaged on 
the source area (second panel). However, also in this case, planetary 
deviations, as measured by the residuals (bottom panel), still are present.
The planetary event MOA-2007-BLG-400 shows a similar shape of the light
curve \cite{dong}.

An example of II class event is given in Fig. \ref{proceeding4} 
where we see that also a small mass planet ($M_{\rm P} = 0.3 ~ M_{\oplus}$) 
can cause detectable planetary deviations in events for which 
the finite size effects are small (see also Fig. 8 in paper \cite{pacz96};
we remind that we have local light minima in the cases 
$d_{\rm P} < R_{\rm E}$).  
In the considered event the geometry is such that the source 
trajectory is passing (in the lens plane) in the region
between the two planetary caustics,
where a deficit of magnification is present.
Other examples of light curves for I and II class events are given in 
\cite{mnras}.

\begin{figure}
\includegraphics[width=0.75\textwidth]{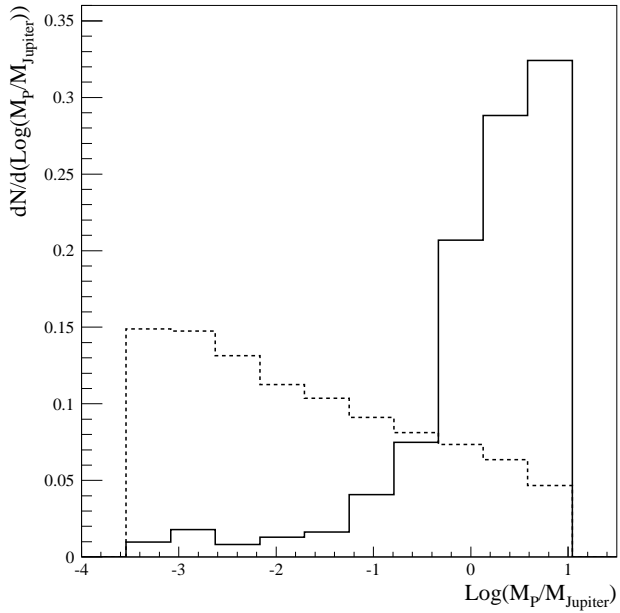}
\caption{Normalised (to unity)  distributions of the exoplanet mass 
$M_{\rm P}$ 
for the events with detectable planetary deviations (solid line) and 
for the generated events (dashed line).}
\label{proceeding5}       % Give a unique label
\end{figure}

\begin{figure}
\includegraphics[width=0.75\textwidth]{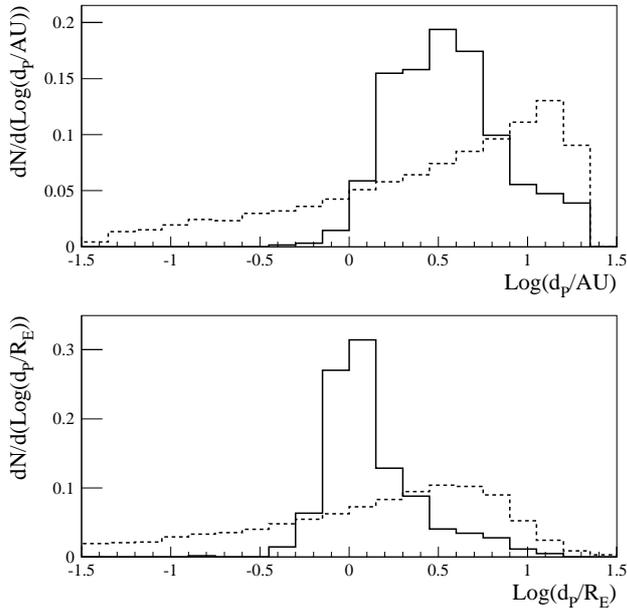}
\caption{Upper panel: normalised (to unity) 
 distributions of the star-to-planet 
separation $d_{\rm P}$ (in AU units)  for the events with detectable planetary 
deviations (solid line) and for the generated events (dashed line).
Bottom panel: distribution of $d_{\rm P}/R_{\rm E}$ 
for events as before.}
\label{proceeding6}       % Give a unique label
\end{figure}

\begin{figure}
\includegraphics[width=0.75\textwidth]{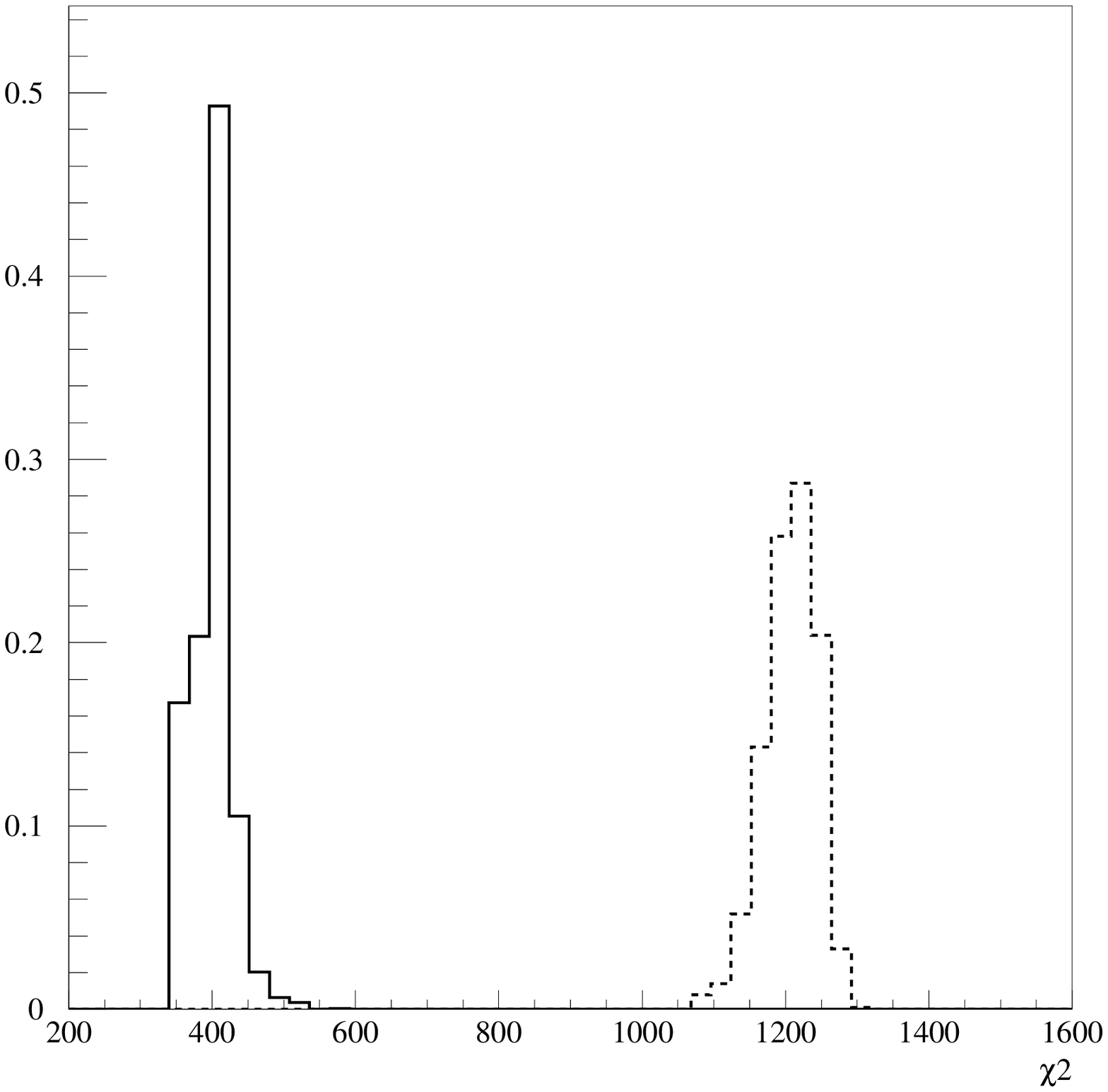}
\caption{Normalised (to unity)  distributions of the  
$\chi^2$ for binary (solid line) and single lens fit 
(dashed line).}
\label{chi2plot}  
\end{figure}

The distributions of the planet mass $M_{\rm P}$  
for the planetary selected events 
($\chi_{\rm r}^2 > 4$, $N_{\rm good}>3$ 
and $\langle \epsilon \rangle_{\rm max} > 0.1$)
is given in the Fig. \ref{proceeding5} (solid line). 
For comparison, the $M_{\rm P}$ distribution for the whole sample of 
detectable events (dashed line) is also given.
From Fig. \ref{proceeding5} it follows that larger planetary masses lead to 
higher probability for the detection of planetary features. 
This result reflects the fact that the detection probability is
proportional to the caustic area (for small size sources), 
which increases with the
planet-to-star mass ratio. 
From the same figure, it also follows that the exoplanet detection 
can occur with a non negligible probability 
for $M_{\rm P} > 0.06~M_{\rm {J}}$ ($M_{\rm P} > 20~M_{\oplus}$),
although even Earth mass exoplanets might be in principle detectable. 
However, if we consider telescopes with smaller diameter ($D<4$ m),
the tail at low masses in Fig. \ref{proceeding5} disappears and, 
practically no exoplanet detection occurs 
for $M_{\rm P} < 0.06~M_{\rm {J}}$.

The probability of planet detection is maximised when the planet-to-star 
separation $d_{\rm P}$ 
is inside the so called ``lensing zone'', 
which is the range of star-to-planet separation 
$0.6   <  d_{\rm P}/R_{\rm E} < 1.6$
\cite{Griest98,gould92}. The $d_{\rm P}$ distribution for selected 
(solid line) and detectable (dashed line) events are shown in the upper panel 
of Fig.~\ref{proceeding6}. 
The relevance of the lensing zone is clarified in the 
bottom panel of the same figure where the planet separation (in unit of the 
Einstein radius) $d=d_{\rm P}/R_{\rm E}$  is plotted.

Our analysis shows that in pixel-lensing searches towards M31 the typical  
duration of planetary perturbations is about 1.4 days. 
However, the number of significant planetary deviations on each light 
curve increases with increasing ratios $\rho/u_0$. So, the 
overall time scale for planetary deviations can  
increase up to 3 -- 4 days (for I class events).
This means that a reasonable value of the time step for pixel-lensing 
observations aiming to detect planets in M31 is a few (4--6)  hours, 
almost irrespectively on the telescope diameter $D$.

We also verify that the overall probability (we stress that here we are
considering the full planetary mass range from 0.1 $M_{\oplus}$ until 10 
$M_{\rm {J}}$,
as well as the  $R_{\rm max}$ and $t_{1/2}$ ranges)
to find pixel-lensing events with
detectable planetary deviations is, however, very small:
less than 2~\% for $D=8$ m and about 3\%, if in the selection procedure
we relax condition ({\it {iii}}) 
(namely, $\langle \epsilon \rangle_{\rm max} > 0.1$) 
and decreases rapidly for smaller telescopes.

%Since perhaps the assumption that all stars have planets
%is too optimistic, more realistically one should further 
%divide these values by at least a factor of two.

\section{Analysis of the PA-99-N2 event case}

The POINT-AGAPE collaboration reported the detection of the microlensing event
PA-99-N2 \cite{paulin03}. This appears as a peculiar event first because of 
its extreme brightness ($R_{\rm max} \simeq 19$ mag), long duration 
($t_{1/2} \simeq 24$ day)
\footnote{As discussed in Section 3, for the II class events
the mean values of $R_{\rm max}$ and $t_{1/2}$ are about 
$23.1$ mag and $6.4$ day, respectively.} and location, some 22 arcmin
away from the M31 center;
second, as shown in a subsequent analysis \cite{an04}, 
because the anomaly with respect to the Paczy\`nki shape along the
light curve can be attributed to a secondary component orbiting the lens star.
In particular, An et al. \cite{an04} have evaluated the 
{\it a posteriori} probability distribution for the lens mass which 
results to be extremely broad: for source and lens disc objects, 
they report a lens mass range 0.02 -- 3.6 $M_{\odot}$ at 95\% confidence 
level. Together with the small best fit binary 
lens mass ratio, $q \simeq 1.2 \times 10^{-2}$, this puts the lens companion
well in the substellar range. In paper \cite{mnras}
we have remarked that taking at face value the most likely value for the 
lens mass $\simeq 0.5~M_{\odot}$ for a disc lens, the lens companion would be 
a $\simeq 6.34~M_{\rm J}$ object. 
This would make of the PA-99-N2 lens companion
the first exoplanet discovered in M31. Furthermore, we had analysed 
the PA-99-N2 event within the framework of our simulation scheme,
showing in particular that its (microlensing and planetary) parameters 
nicely fall in the expected range for II class events \cite{mnras}. 
Here we further analyse this event. First, starting from the observational data
\footnote{Courtesy of the POINT-AGAPE collaboration.},
we test the robustness of the binary-lens best fit
solution.  Second, we address the question 
of the efficiency for finding binary-like
deviations for such bright and long duration events.

In Table 1 of An et al. \cite{an04} a list of the best fit parameters adopting
various models, including the best fit assuming a single lens, for different 
location of the lens system are given. In particular, the two best fit
binary models, named C1 and W1, are indistinguishable
\footnote{There exist several binary lens models that lie at a local
$\chi^2$ minimum. Some of these degeneracy would have been removed  if 
the $\simeq 20$ day observation gap between JD' 72 and JD' 91 had been 
regularly covered \cite{an04}.}  in terms of their $\chi^2$.
{ An et al. \cite{an04} considered finite source corrections for the C1 
model (they named this model as FS) and found that these corrections decrease 
slightly the $\chi^2$ ($\simeq 2\%$) with respect to C1 model.
Therefore, these authors claimed that C1 and FS models are practically 
the same. In our analysis we consider the  C1 model. To verify  
the robustness of the binary-lens fit solution, 
the first test is to add a Gaussian noise to the best fit of the observed 
light curve to verify if a single lens model with noise can reproduce the 
observational data. We also let each parameter vary (by at most 20\%) around 
the given central value. For each of the so obtained light curve we 
calculate the $\chi^2$,
which is plotted in Fig. \ref{chi2plot} (dashed line). We have verified
that the lowest $\chi^2$ value corresponds to the parameters of the 
best fit Paczy\'nski model (last row) in Table 1 of An et al. \cite{an04}.
Similarly, we have taken the best fit binary model C1  
and realised more than one thousand models by adding
Gaussian noise and letting the parameters to vary by at most 20\%.
We plot again in Fig. \ref{chi2plot} the $\chi^2$ distribution
(solid line). Also here the lowest $\chi^2$ value is obtained for the C1 
model without noise. 
Moreover, from Fig. \ref{chi2plot} one can see that the two
(normalized) distributions of the $\chi^2$ 
are clearly separated, which implies that
the best single lens fit is much worst than any of the binary lens
models. From this we conclude that the binary fit is robust and that
the observed light curve cannot be obtained by any single lens model
with random noise.

A further point to be stressed is the following: at the end of the previous 
section we have mentioned  that the probability 
to detect an  exoplanet in pixel-lensing observations
towards M31 is rather small even with large telescopes.
Therefore, the question which arises now is the following: 
what is the chance of finding a planetary feature in an event as PA-99-N2? 
The basic answer can be found looking at the characteristics
of the underlying microlensing event. As it is also apparent by comparing 
with the distributions shown in Fig. \ref{proceeding1}, 
this is, at the same time, an extremely
bright and long duration event. In fact, as we now show,
this strongly enhances the probability for finding
deviations to the single lens shape. 
To this purpose we perform a MC simulation
where we fix the single lens parameters $R_{\rm max}$ and $t_{1/2}$ 
to those of PA-99-N2 and we let vary the binary ones
(planet-to-star mass ratio, planetary distance in unit of $R_{\rm E}$ and 
orientation of the source trajectory with respect to the binary axis) 
for an observational setup fixed to reproduce
that of the POINT-AGAPE observations
(mirror size, sampling and exposure time). 
In particular we find, for the selected planetary mass range, an increase
of the average efficiency up to 6\%, to be compared with 
less than 0.6\% (see Table 3 in \cite{mnras}) 
for events without any constraints on $R_{\rm max}$ and $t_{1/2}$. 
Moreover, as also implicit from the analysis of Fig. \ref{proceeding5}, 
the efficiency for finding binary-like deviations in the simulated light 
curves increases significantly with the lens-companion mass.
In particular, for the restricted range $1-10 M_{\rm J}$ we find
the average efficiency rises up to 27\%.
We have restricted our attention to the planetary-mass range
for the lens companion ($M_{\rm P} < 10 ~M_{\rm J}$ 
was our initial assumption),
but we might expect the efficiency still rises
once we enter the brown dwarf mass range
(as in fact it is allowed by the analysis \cite{an04}). 
Finally,  the efficiency would be further enhanced for a
lens-companion mass function more peaked towards higher values
of the mass than that we have assumed.

\begin{figure}
%\includegraphics[width=0.75\textwidth]{n2-bin.eps}
\includegraphics[width=1.\textwidth]{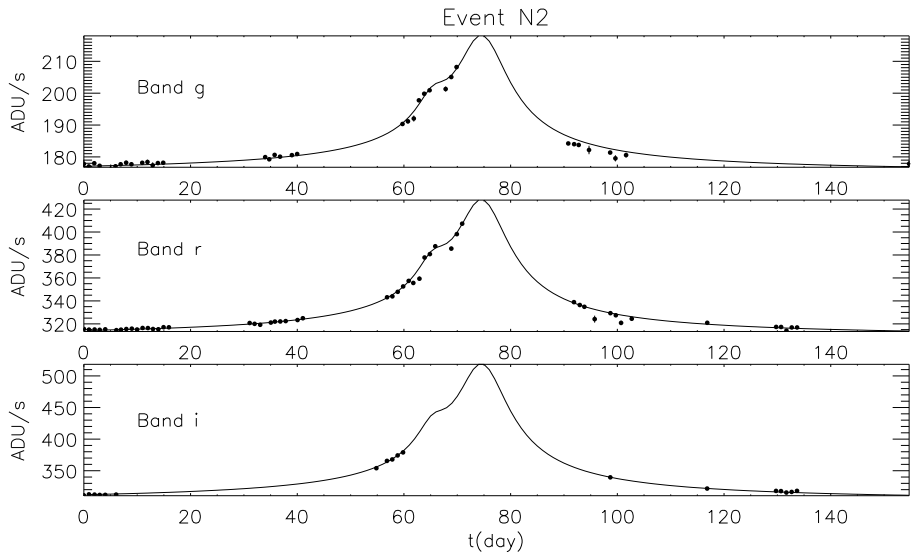}
\caption{The binary light curves corresponding to the C1 model
\cite{an04} for g, r and  i bands.}
\label{n2-bin.eps}       % Give a unique label
\end{figure}

\section{Conclusions}

We have discussed the possibility to detect planets in M31 by using
pixel-lensing observations with telescopes of different sizes and
observational strategies. 
Assuming that each lens star is hosting a planet,
we carry out MC simulations and explore the multi-dimensional space of the 
parameters for both the lensing and planetary systems.
Planetary induced perturbations in light curves are detected by comparing 
the simulated light curves with the corresponding Paczy\'nski shapes 
and searching for significant deviations.
The MC approach allows to characterise the sample of events for which the 
planet detections are more likely to be observed.
Since in pixel-lensing towards M31 the bulk of the source stars are
red giant (with large radii), we take into account
the finite source effects, which
induce a smoothing of the planetary deviations 
and decrease thus the chance to detect planets.
We estimate the typical duration of a single planetary feature to be of 
about one day. However, the number of significant
planetary deviations and 
consequently the overall time scale of the perturbations increases 
(up to a few days), increasing the source size. 
Therefore for pixel-lensing searches towards M31 only few 
exposures per day could be 
enough to detect planetary features in light curves. 
Pixel-lensing technique favours the detection of large mass planets 
($M_{\rm P} \simeq 2~M_{\rm {J}}$), even if planets with mass less than 
$20~M_{\oplus}$ could be detected, although  with small probability, 
by using large telescopes with a sufficient photometric stability.

As a test case for exoplanetary searches in M31
we have reconsidered the POINT-AGAPE event
PA-99-N2, which had already been probed to show
an anomaly with respect to the Paczy\`nski shape
compatible with a binary lens. As a first
step we have revisited the issue of 
the single lens versus binary lens solution,
finding that the latter is indeed robust
against the introduction of a gaussian noise
along the observed data. According to the previous
POINT-AGAPE analysis this binary system has
a small mass ratio and this makes
at least plausible that the lens companion
is indeed an exoplanet. Furthermore, 
the underlying microlensing event is
extremely, and somewhat unusually, long and bright.
Therefore, as a second step, we have
carried out a specific MC
simulation that allowed us to show
that for this kind of events the chance of finding
exoplanetary deviations is indeed greatly enhanced and
possible even for an observational set up
as that of the POINT-AGAPE observations. As a caveat, we mention that 
the efficiency grows with the lens companion mass
also beyond the exoplanet mass range in the brown dwarf one.

Whatever the case for PA-99-N2 event, our analysis
confirms that looking for exoplanets in M31 with 
pixel-lensing, at least in the Jupiter mass range,
is already reachable with  present technology. 
Clearly, an efficient strategy of search,
as towards the Galactic bulge, is
mandatory: a wide field survey,
to collect a large enough number
of pixel-lensing candidates, endowed with
an early warning system to trigger
subsequent follow up observations, possibly with a network
of telescopes around the world (for which also telescopes 
with small field  of view could be usefully employed). 
In our opinion the reward of such a project would be substantial: going from 
the settling of the question of the MACHO fraction in M31 halo, to important 
information on the stellar mass function and the detection of exoplanets, 
besides other information on the M31 structure and  content of variable stars.

%{\it 
%Finally we note that gravitational microlensing is a very efficient
%method for discovering exoplanets around habitable zone, namely 
%planetary systems with Earth-like masses and distances of about AU. 
%In this respect, besides ongoing and foreseen observational programs
%\cite{dominik,beaulieu}, Microlensing Planet Finder (MPF) mission 
%would be very fruitful in comparison with other space missions for exoplanet
%searches (see Fig. 2 in \cite{Bennett_08b} and  Fig. 1.9 in
%\cite{bennett09}). 
%}

\begin{acknowledgements}
SCN    acknowledges support for this work
by the Italian Space Agency (ASI)
and by the ``Istituto Internazionale per gli
Alti Studi Scientifici'' (IIASS).
AFZ thanks a financial support by INFN (Sezione di Lecce) 
at Dipartimento di Fisica (Salento University).
\end{acknowledgements}

\end{document}